\documentclass[proof]{WileyASNA-v1}

\articletype{Article Type}%

\received{01 February  2023}
\revised{01 June 2023}
\accepted{14 June 2023}
\doiheadtext{10.1002/asna.20230103}

\graphicspath{{figures/}{./}}

\usepackage[euler]{textgreek}
\def\um{{\textmu}m}

\def\elec{$\mathrm{e^-}$}

\def\invf{$1/f$}
\def\dshrng{\,$-$\,}
\def\spaceK{\,K}

\hypersetup{colorlinks=true,allcolors=blue}
\usepackage{cleveref}

\Crefformat{figure}{Figure~#2#1#3\hspace{-5pt}}
\Crefformat{table}{Table~#2#1#3\hspace{-5pt}}

\usepackage{enumitem}

\begin{document}

\title{Evaluating the GeoSnap 13-\um\ Cut-Off HgCdTe Detector for mid-IR ground-based astronomy}

\author[1]{Jarron M.\ Leisenring*}
\author[2]{Dani Atkinson}
\author[2]{Rory Bowens}
\author[3]{Vincent Douence}
\author[1]{William F.\ Hoffmann}
\author[2]{Michael R.\ Meyer}
\author[3]{John Auyeung}
\author[3]{James Beletic}
\author[4,5]{Mario S.\ Cabrera}
\author[6]{Alexandra Z.\ Greenbaum}
\author[7]{Phil Hinz}
\author[8]{Derek Ives}
\author[4]{William J.\ Forrest}
\author[4]{Craig W.\ McMurtry}
\author[4,9]{Judith L. Pipher}
\author[2]{Eric Viges}

\authormark{Leisenring \textsc{et al}}

\address[1]{
  \orgdiv{Steward Observatory and Department of Astronomy}, 
  \orgname{University of Arizona}, 
  \orgaddress{933 N Cherry Ave, Tucson, \state{AZ}, 85721 \country{USA}}}
\address[2]{
  \orgdiv{Dept of Astronomy}, 
  \orgname{University of Michigan}, 
  \orgaddress{1085 S University Ave, Ann Arbor, \state{MI}, 48103 \country{USA}}}
\address[3]{
  \orgname{Teledyne Imaging Systems}, 
  \orgaddress{5212 Verdugo Way, Camarillo, \state{CA}, 90312 \country{USA}}}
\address[4]{
  \orgdiv{Dept of Physics \& Astronomy}, 
  \orgname{University of Rochester}, 
  \orgaddress{206 Baucsh and Lomb Hall, Rochester, \state{NY}, 14627 \country{USA}}}
\address[5]{
  \orgname{Conceptual Analytics LLC}, 
  \orgaddress{8209 Woburn Abbey Rd, Glenn Dale \state{MD}, 20769 \country{USA}}}
\address[6]{
  \orgdiv{IPAC}, 
  \orgname{CalTech}, 
  \orgaddress{1200 E. California Blvd., Pasadena, \state{CA}, 91125 \country{USA}}}
\address[7]{
  \orgdiv{Astronomy \& Astrophysics Department}, 
  \orgname{UC Santa Cruz}, 
  \orgaddress{1156 High St., Santa Cruz, \state{CA}, 95064 \country{USA}}}
\address[8]{
  \orgdiv{Org Div},
  \orgname{ESO},
  \orgaddress{Karl-Schwarzschild-Strasse 2, Garching, \country{Germany}}}
\address[9]{Deceased.}

\corres{*\email{jarronl@arizona.edu}}

\presentaddress{933 N. Cherry Ave, Tucson, AZ 85721, USA}


\abstract{
New mid-infrared HgCdTe (MCT) detector arrays developed in collaboration with Teledyne Imaging Sensors (TIS) have paved the way for improved 10-\um\ sensors for space- and ground-based observatories. 
Building on the successful development of longwave HAWAII-2RGs for space missions such as NEO Surveyor, we characterize the first 13-\um\ GeoSnap detector manufactured to overcome the challenges of high background rates inherent in ground-based mid-IR astronomy.
This test device merges the longwave HgCdTe photosensitive material with Teledyne’s 2048$\times$2048 GeoSnap-18 (18-\um\ pixel) focal plane module, which is equipped with a capacitive transimpedance amplifier (CTIA) readout circuit paired with an onboard 14-bit analog-to-digital converter (ADC). 
The final assembly yields a mid-IR detector with high QE, fast readout ($>$85\,Hz), large well depth ($>$1.2 million electrons), and linear readout. 

Longwave GeoSnap arrays would ideally be deployed on existing ground-based telescopes as well as the next generation of extremely large telescopes. 
While employing advanced adaptive optics (AO) along with state-of-the-art diffraction suppression techniques, instruments utilizing these detectors could attain background- and diffraction-limited imaging at inner working angles $<$10\,\textlambda/D, providing improved contrast-limited performance compared to JWST MIRI while operating at comparable wavelengths. 
We describe the performance characteristics of the 13-\um\ GeoSnap array operating between 38 and 45\spaceK, including quantum efficiency, well depth, linearity, gain, dark current, and frequency-dependent (\invf) noise profile. 
}


\maketitle

\section{Introduction}\label{sec1}

Ground-based mid-infrared (mid-IR) instrumentation has the ability to revolutionize a wide range of astronomical research, but presents significant challenges.  
Emission from 3\dshrng13\,\um\ 
can be used to characterize astrophysical objects over a wide range of temperatures ($<$100 to $>$1000 Kelvins), penetrate obscuring dust, investigate physical conditions of environments containing molecules bearing key volatile elements of carbon, nitrogen, and oxygen, and even probe galaxy formation and evolution in the early Universe.  However, ground-based mid-IR observing has sometimes been compared to observing in the visible during the daytime with a telescope that is on fire, because all terrestrial objects emit significantly at these wavelengths, providing an overwhelming high background.  The development of mid-IR photo-conductive detectors began in the 1960s and progressed rapidly.  While many pioneering observations were made with ground-based systems, it was clear that mid-IR astronomy in space would have tremendous returns leading to the launch of IRAS in the 1980s \citep{neug1984}.

For diffraction-limited observations of point sources, the time to complete an observation at a fixed signal-to-noise (SNR) scales as $D_{\rm tel}^{-4}$ when the dominant source of noise is background radiation.  This was achievable from the ground on 4-meter class telescopes in the 1990s as long as a bright source remained in the field of view throughout the observing sequence to serve as a reference.  Lacking such sources, it was difficult to keep the telescope fixed and integrate for long periods on the same field to reveal faint targets.  The advent of tip-tilt secondaries and adaptive optics (AO) led to huge improvements \citep[e.g.;][]{beck1993,davi2012}.  However, detector limitations such as low quantum efficiency, as well as excess low frequency noise which required modulating the source on the detector at high frequency (chopping), limited what could be done.  

The launch of NASA's \textit{Spitzer Space Telescope} in 2003 provided an extremely sensitive and reliable platform for wide-field mid-IR observations, enabling extraordinary progress, albeit with limited spatial resolution.  
With the more recent launch of the 6.5-meter JWST \citep{gard2006}, why should we invest now in ground-based mid-IR? JWST MIRI enables a tremendous leap in mid-IR astronomy with a large field of view and extraordinary sensitivity \citep{riek2015, wrig2015}. Yet, due to the mid-frequency errors in the mirror segments and limited stability and metrology, JWST has limited contrast as a function of $\lambda/D$, as demonstrated in \cite{cart2022}. Instead, ground-based telescopes equipped with advanced AO will outperform JWST in the contrast limit around bright sources \citep[cf.;][]{dani2018, bocc2015, dela2013, beic2019,ren2023}. 

There are many scientific areas where a sensitive high resolution/contrast mid-IR capability can provide fundamental breakthroughs including:  a) exoplanet detection and characterization; b) star formation and planet-forming disks; c) circumstellar environments of evolved stars; d) warm dust emission in local group star clusters; e) star forming galaxies and active galactic nuclei in the local Universe; and e) multiwavelength observations of multiple strong gravitational lens sources at high redshift.  Ground-based AO assisted mid-IR imaging is extremely complementary to what JWST can provide.  

Currently, there are only two mid-IR detectors on large ground-based telescopes equipped with adaptive optics that have recently been in operation.  The NEAR experiment on the ESO VLT \citep{kasp2017} was funded by Breakthrough Initiatives to search for Earth-like planets around the nearest stars to the Sun, $\alpha$ Cen A and B.  The system utilizes a Raytheon AQUARIUS array with modest quantum efficiency ($\sim$ 40\%) and excess low frequency noise \citep[e.g.;][]{hoff2014}.  Despite these limitations, the NEAR project produced interesting results \citep{kasp2019, wagn2021, path2021, visw2021}, but is no longer available.  Another system, NOMIC on the LBTI, also utilizes an AQUARIUS array, with the vast majority of its observations obtained with nulling interferometry \citep[e.g.;][]{erte2018}.

Teledyne Imaging Sensors (TIS) developed a longwave ($<$15\,\um) large format (2048 $\times$ 2048) HgCdTe (MCT) detector for low background space-based applications with high quantum efficiency (approaching 90\%) and low noise \citep{mcmu2013}.  A parallel effort has produced the GeoSnap array, with deep wells and a fast read read-out integrated circuit (ROIC) potentially suitable for ground-based astronomy.  This new ROIC paired with mid-infrared material could revolutionize ground-based, mid-IR astronomy, particularly when coupled with adaptive optics on 6\dshrng12-meter telescopes.  Such systems on 25\dshrng39-meter class telescopes would be capable of imaging rocky planets around the nearest stars \citep[e.g.;][]{hinz2010, bowe2021}.  Here we describe efforts to characterize the GeoSnap array for use in ground-based mid-IR instruments.  One such instrument is MIRAC-5, currently being refurbished for use on the MMT with the new MAPS AO system \citep{bowe2022,morz2020}.  

We describe the GeoSnap device in \Cref{sec:device} and provide an overview of the test facilities in \Cref{sec:test_stations}. The device's measured characteristics are presented in \Cref{sec:results}, while \Cref{sec:pink_noise} separately discusses observed excess noise.  

\section{Device Description}
\label{sec:device}

\begin{figure}[!h]
  \begin{center}
    \includegraphics[width=85mm]{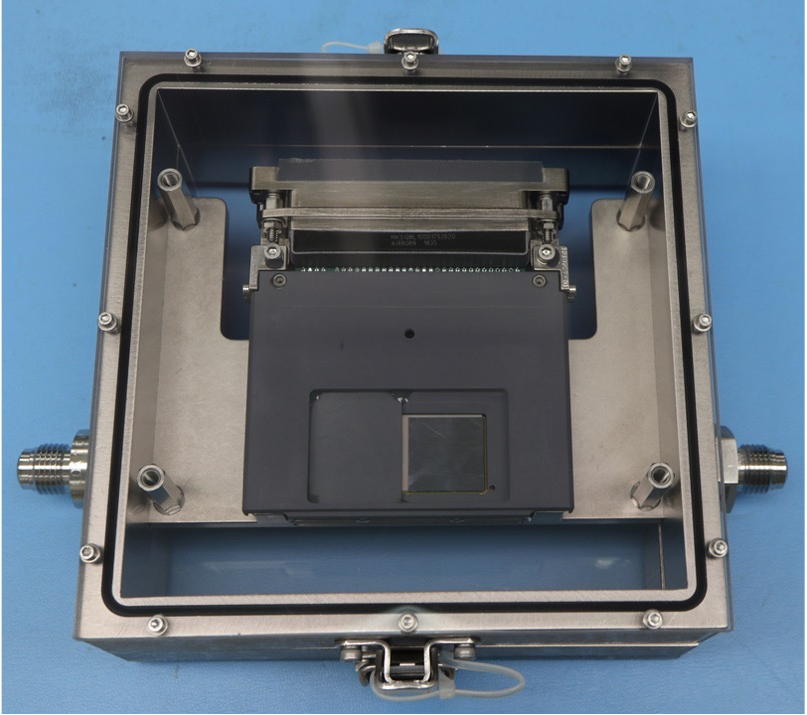}
  \end{center}
  \caption{The GeoSnap-18 (\texttt{A0} ROIC) Part Number SN20561 enclosed in its storage case. The hybridized quadrant is evident in the lower right corner, while the three remaining quadrants are covered with a mask. Two layers of ribbon cables connect to the top orientation, carrying power and digital lines.}
  \label{fig:geosnap_part}
\end{figure}

The GeoSnap-18 is a large-format (2048$\times$2048) detector array with an 18-\um\ pixel pitch developed by Teledyne Imaging Sensors (TIS) for fast imaging operations. The standard full frame readout maxes out at 86.7\,Hz frame rate, but has capabilities for up to 143.2\,Hz. The readout integrated circuit (ROIC) utilizes a capacitive transimpedance amplifier (CTIA), which enables large well depths (1.2\dshrng2.6\,M\elec and higher) relative to the traditional pixel source follower ($\sim$100\,k\elec) found in the HAWAII-xRG (HxRG) detector series \citep{bele2008, jerr2019}. 

GeoSnap incorporates on-chip electronics for clock timing, bias generation, and 14-bit ADC within a single assembly located at the instrument focal plane. 
All pixels are exposed simultaneously using a global shutter, differing from standard HxRG operations, which utilize a rolling clocking scheme such that consecutive pixels are addressed at slightly different times (e.g., 10-{\textmu}sec increments). 
The chip incorporates an integrate while read architecture, which minimizes overheads and maximizes exposure time efficiency. 
GeoSnap's standard full frame readout can operate below 1\,Hz and up to 86.7\,Hz frame rate, although frame rates of $>$140\,Hz can be achieved at the cost of reduced ADC resolution (13.2 bits). Vertical window mode enables higher frame rates when reading $<$2048 rows.

This combination of high speed and large well depths make these arrays suitable for high-background IR environments.
\Cref{fig:geosnap_part} shows an image of the GeoSnap focal plane module SN20561, consisting of a 13-\um\ cut-off HgCdTe die hybridized onto a single 1024$\times$1024 quadrant. The HgCdTe material originated from a program led by the University of Rochester to develop longwave HxRG arrays out to 15\,\um\ \citep[e.g.,][]{mcmu2016,dorn2018,cabr2019,cabr2020}. 

\subsection{Unit Cell}
\label{sec:unitcell}

For a typical source follower readout used in the HxRG detectors, photoelectrons travel primarily via diffusion while under the influence of a weak electric field within the photosensitive material \citep{pain1993, riek2002}. The output signal is simply a measure of the voltage strength across the HgCdTe p-n junction as photo-generated charge travels across the depletion region. As more electrons are knocked loose, the strength of the electric field drops, which naturally limits well sizes to approximately 100\,k\elec\ depending on the initial configuration of the device's backbias.  

\begin{figure}[!h]
  \begin{center}
    \includegraphics[width=85mm]{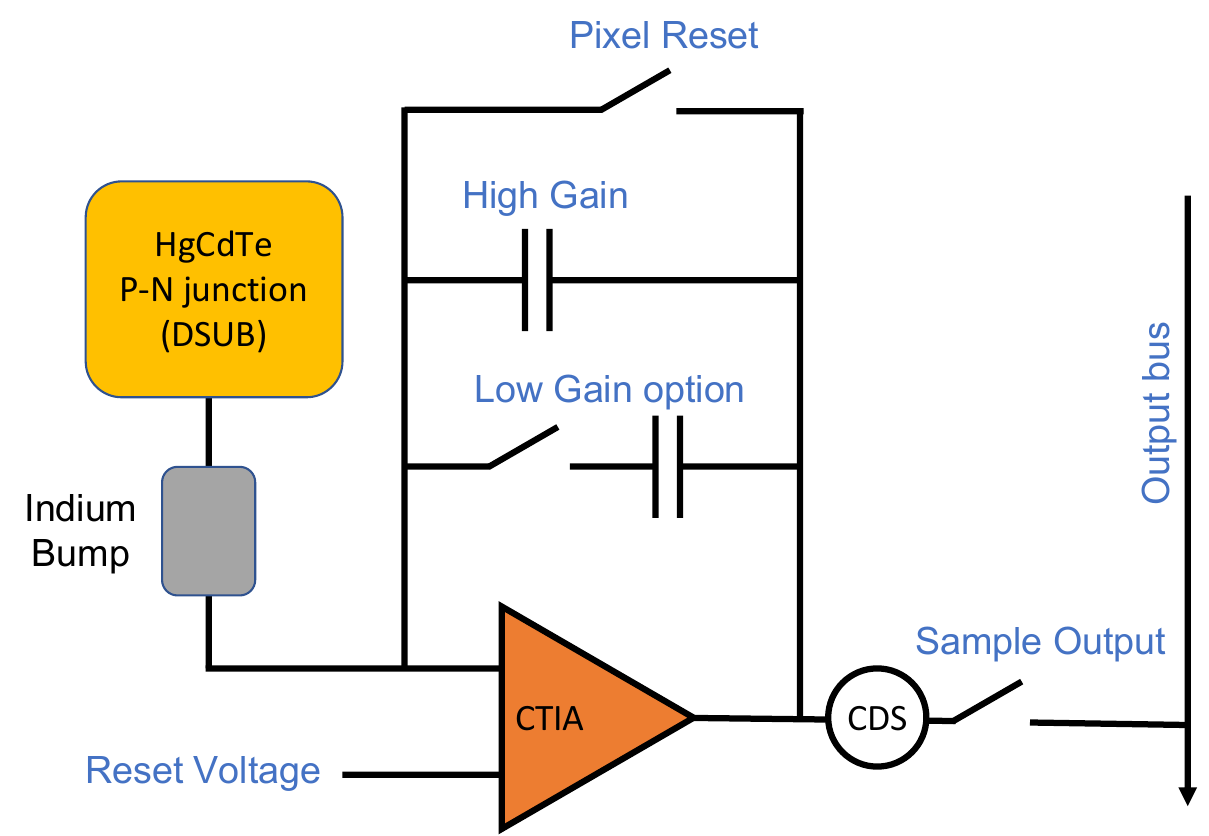}
  \end{center}
  \caption{Cartoon schematic of CTIA pixel unit cell design.}\label{fig:ctia}
\end{figure}

For a CTIA circuit as depicted in \Cref{fig:ctia}, the input gate of the HgCdTe material is constantly held at the reset voltage during detector operations. The signal is then allowed to accumulate on a capacitor embedded within each pixel unit cell. This readout scheme enables large well depths defined by the capacitors with highly linear outputs. 

The GeoSnap multiplexer includes two different gain modes using independent capacitors with different inherent charge-well sizes. Well sizes vary depending on ROIC revision; the \texttt{A0} version provided for this study was manufactured for nominal well depths of 1.2\,M\elec\ and 90\,k\elec\ for the low and high gain modes, respectively. The more recent \texttt{A1} and \texttt{B0} ROICs have approximately twice the well depths for both gain modes (2.7\,M\elec\ and 190\,k\elec\ for the low and high gain modes, respectively). Read noise in the low gain mode typically runs 5-7 times higher that of high gain (ie., 150\,\elec\ versus 25\,\elec).

Within an observed frame time, GeoSnap's CTIA circuit performs a number of operations prior to digitization of the signal, including a pixel reset, on-chip CDS, and integration. At the beginning of a frame, the programmable reset duration holds the pixel reset switch closed, preventing signal accumulation on the capacitors. The reset time is configured as a fraction of the total frame time, and integration is simply the remaining frame time minus small internal switching overheads. Prior to integration, the circuit quickly samples and stores the reset state to perform an analog on-chip CDS subtraction of the measured signal. More complex readout modes have been developed to disable the reset and CDS operations within a subsequent number of frames in order to perform a sample-up-the-ramp integration with multiple frames. 

The CTIA circuit can also eliminate some problematic detector characteristics that plague observations with HxRG detectors. Since the depletion region of the pixel P-N junction is held constant throughout GeoSnap operations, charge traps remain unfilled since no de-biasing occurs. This effectively removes charge persistence effects and related latent images. In addition, holding all pixel gates at the same constant reset voltage eliminates interpixel capacitance (IPC) caused by capacitive coupling of neighboring pixels \citep{moor2004}. The constant voltage fields also prevent the ``brighter-fatter'' phenomenon where a PSF appears to increase in size with brightness due to charge migration from bright pixels to their neighbors during an exposure \citep{plaz2018}.

Power dissipation at the focal plane is extremely high due to the CTIA circuit being held at a constant reset voltage along with the high-speed readout and onboard digitization. The resulting heat leads to significant ROIC glow that is easily detectable by longwave devices. However, for ground-based mid-IR observations, the background rates from the sky and warm telescope optics dominate over the GeoSnap's self-emission.


\subsection{Readout Architecture}

The readout electronics used in this study consist of three primary components: the cryogenic Focal Plane Module (FPM), external Focal Plane Electronics (FPE), and Processing Electronics (PE) board. 
These are connected in series with various cable harnesses to facilitate communication interfaces and science data transfer. Each of the component contains test pattern modes to generate predictable images to check the electronics' functionality and verify connectivity.

The FPM is comprised of the ROIC, an on-chip 14-bit ADC, bias generation, and a programmable serial interface. 
Digitized pixel data is converted to 16-bit information and sent through 8 digital output ports mapped to 512$\times$1024 sections of pixels. The image output is passed to the FPE through two ribbon cable harnesses at 1.6\,Gbps per output, for a total bandwidth of 12.8\,Gbps. 
In addition to receiving and processing the science pixel data, the FPE provides the power, control, and clock signals to operate the ROIC. 

The PE sits between the FPE and host computer, primarily acting as a control interface to the FPE and ROIC as well as converting image data to the Camera Link interface format for ingestion by the host computer. User commands are sent over a USB cable using a standard serial connection. Video data is received by the host computer over four Camera Link cables, each one dedicated to an independent detector quadrant. Data is received by a Matrox Radient ev-CL PCIe frame grabber capable of handling the full 2048$\times$2048 16-bit data at $>$85\,Hz without overheads. We have implemented a Linux driver to continuously process and write contiguous frames in real time.

\subsection{Low Operating Temperatures}

At operating temperatures below 55K, the master voltage reference of the early generation \texttt{A0} and \texttt{A1} ROICs is unable to attain its nominal value of 3.0\,V, and instead rails at 3.3\,V no matter the register setting of the programmable DAC. This is because these ROICs were designed for nominal operations of $>$77\spaceK. At lower temperatures, the built-in bandgap reference that defines the settable DAC range of the master voltage would substantially increase. The more recent \texttt{B0} ROIC revision includes an option to use non-bandgap derived master reference for this range of temperatures.

While all data presented in this work were acquired using an \texttt{A0} ROIC operating between 35K and 50\spaceK\ to reduce dark current, preliminary results did not show any deleterious effects to the data when compared to observations acquired at T=80\spaceK. We conclude that a master reference value of 3.3\,V is acceptable for our purposes.

\section{Test Environments}
\label{sec:test_stations}

The GeoSnap-18 hardware (part SN20561) was first delivered to the University of Arizona in March, 2019 for initial check-out and characterization. Primary test data were acquired between August and November of 2019 in a refurbished version of the Mid-InfraRed Array Camera (MIRAC) instrument at Arizona \citep{hoff1993, hoff1998}. 
The detector and readout electronics were subsequently shipped to the University of Michigan for further testing under dark conditions in the MITTEN cryostat \citep{bowe2020}.

\begin{center}
  \begin{table*}[!t]%
  \centering
  \caption{Device Performance Summary\label{tbl:summary}}%
  \tabcolsep=0pt%
  \begin{tabular*}{1\textwidth}{@{\extracolsep\fill}lccl@{\extracolsep\fill}}
  \toprule
  \textbf{Specification} & \textbf{GeoSnap-18} & \textbf{AQUARIUS}\tnote{$\dagger$}  & \textbf{GeoSnap Comments} \\
  \midrule
  Array Size & 2048~$\times$~2048  & 1024~$\times$~1024  & Only 1k~$\times$~1k active; available in 1k, 2k, or 3k formats \\
  Pixel Size & 18\,\um  & 30\,\um  & \\
  Operating Temperature & 35\,--\,50\spaceK\  & 7\dshrng9\spaceK\  & Absolute lower limit not characterized in this study \\
  Max Full Frame Rate & 86.7\,Hz& 100\,Hz & 120\dshrng140\,Hz in special configuration \\
  Wavelength Range & 1\dshrng13\,\um\ & 3\dshrng28\,\um\ & Can range 0.4\dshrng15\,\um, depending on HgCdTe bandgap \\
  Quantum Efficiency & >70\% & >40\% & >90\% with thinned CdZnTe and AR coating \\
  Well Depth & 1.3\,M\elec\ & 0.8\dshrng1\,M\elec\ & >2.5\,M\elec\ for \texttt{B0} revision; 180\,k\elec\ in high gain mode \\
  Non-linearity & <5\% & <5\% & <0.1\% up to full well after correction \\
  Dark Current (per pixel) & >10,000 \elec/sec & <100\,\elec/sec & Measured at 38\,K; limited by ROIC glow \\
  Dark Current Density & >0.5\,nA/cm$^2$ & <0.002\,nA/cm$^2$ &  \\
  Read Noise & 140\,\elec\ & 200\,\elec\ & $\sim$90\%\ of kTC removed within unit cell processing \\
  Excess Noise & \invf\ & ELFN & Mitigated through chopping and slower readout speed \\
  Power Dissipation & 1000\,mW & 250\,mW & 250\dshrng500\,mW when operating only a single quadrant \\
  \bottomrule
  \end{tabular*}
  \begin{tablenotes}
    \item GeoSnap performance summary as measured in low gain mode and comparison to AQUARIUS characteristics
    \item[$\dagger$] \cite{ives2012, ives2014, hoff2014}
  \end{tablenotes}
  \end{table*}
\end{center}

\begin{figure*}[!hb]
  \begin{center}
    \includegraphics[width=1\textwidth]{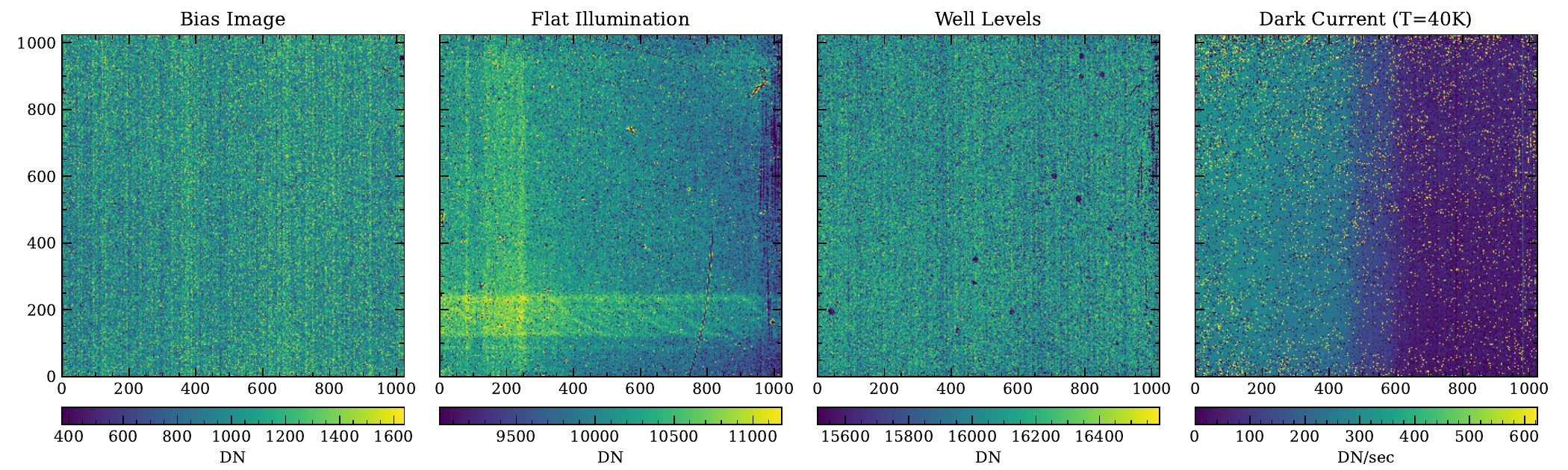}
  \end{center}
  \caption{Representative images of the array image properties, including bias frame, flat field illumination, well level distribution, and dark current. The ``plaid'' stripe pattern in the Flat Illumination panel is consistently observed in different cryostats when the array is subjected to sufficient illumination, likely due to reflections off of the bezel edges surrounding the hybridized region. The gradient in the dark current image likely relates to MUX glow.
  \label{fig:overview}}
\end{figure*}

\vspace{-1\baselineskip}

\subsection{MIRAC Testing at AZ}
\label{sec:mirac}

The fifth iteration of the instrument, MIRAC-5's optical design consists of an off-axis ellipsoid to reimage an intermediate focal plane (located near the instrument entrance window) onto the detector surface housed within a cryogenic environment \citep{bowe2022}. 
The entrance aperture wheel has five positions containing three slits ranging from 0.4 to 0.8\,mm wide, a pinhole 0.5\,mm in diameter, and a large square aperture. 
A cold pupil stop is positioned after the reimaging optics. 
Two filter wheels sit immediately after the pupil stop holding over a dozen narrow, medium, and broadband filters throughout the JHKLMN infrared bands as well as blanks to acquire unilluminated observations. 

MIRAC is cooled by a Cryomech Pulse Tube cryocooler, attaining operating temperatures of $\sim$20\spaceK\ for the optics and 30\dshrng50\spaceK\ for the detector.
A thermal leak near the detector focal plane prevented dark current measurements with this instrument. Instead, operations with MIRAC-5 focused primarily on characterizing the detector's linearity, gain, and excess noise properties under illumination.

\subsection{MITTEN Testing at MI}
\label{sec:mitten}

The Michigan Infrared Test Thermal ELT N-band (MITTEN) Cryostat was used to test GeoSnap from the spring of 2020 through the summer of 2021 \citep{bowe2020}. The MITTEN optical bench can reach temperatures $<$20\spaceK\ with interior surfaces reaching $\sim$40\spaceK\ enabled with a two-stage pulse-tube cryocooler package from Cryomech. It has an internal working volume of 0.128\,$\textrm{m}^3$. To minimize external leaks and stray light, MITTEN does not have a window for light to enter from the outside; instead, an internal light source has been installed.
The detector mount is a molybdenum block attached to 6061 aluminum stand mounted on the optics plate. The detector temperature can be controlled on to $\pm$0.001\spaceK\ within 30\dshrng70\spaceK. A minimum temperature of 35\spaceK\ and a maximum temperature changing rate are enforced to protect the detector. 

MITTEN also includes an internal blackbody source, filter wheel, and pinhole point-sources paired with an Offner relay to perform measurements of quantum efficiency and response to point sources relative to thermal background. However, this work focuses primarily on taking advantage of the cryostat's cold and dark interior to measure dark current, ROIC glow, read noise, and confirming the excess noise profiles of the GeoSnap device in a different operating environment.


\section{GeoSnap Characterization}
\label{sec:results}

A single quadrant of the full GeoSnap 2048$\times$2048 ROIC (18-\um\ pixels) was hybridized to a long-wave HgCdTe material with sensitivity extending past $\sim$13\,\um. 
The die material bonded to this device comes from the same wafer as H1RG-18508 characterized in \cite{cabr2019} and should have similar properties, such as cut-off wavelength, quantum efficiency (QE), and intrinsic dark current.

Unless otherwise specified, reported values refer to the \texttt{A0} ROIC operating in low gain mode (large well depths) with an applied detector reverse bias of 130\,mV. \Cref{tbl:summary} summarizes the main features of the detector, while \Cref{fig:overview} highlights representative images for the detector bias, flat field illumination pattern, detector well level, and dark current distributions.

\subsection{Quantum Efficiency}
\label{sec:qe}

\begin{figure}[!t]
  \begin{center}
    \includegraphics[width=87mm]{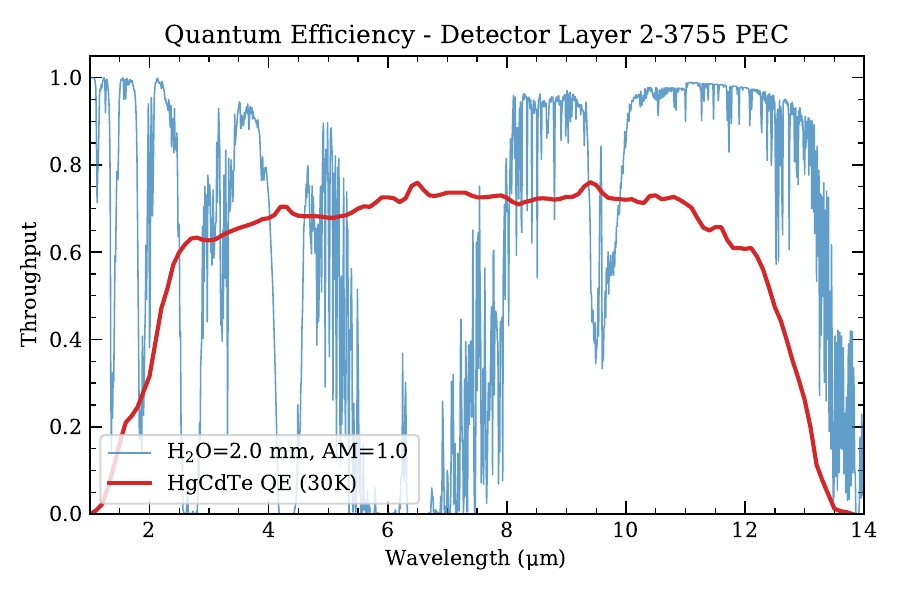}
  \end{center}
  \caption{Quantum efficiency provided by TIS of process evaluation chip (T=30\,K) relative to ATRAN model with 2\,mm of precipitable water vapor at an airmass of 1 \citep{lord1992}. 
  \label{fig:qe}}
\end{figure}

\Cref{fig:qe} shows the wavelength-dependent QE measurements of the process evaluation chip (PEC) at T$=$30\spaceK\ provided by TIS, overplotted with a typical atmospheric transmission curve.
The hybridized quadrant of the GeoSnap device has a wavelength sensitivity ranging approximately between 1 and 13.5\,\um, covering the JHKLMN atmospheric windows. 
The average QE at 10\,\um\ is $\sim$0.7 with a longwave half-power point corresponding to a cut-off of 12.75\,\um. For higher operating temperatures, we expect this cut-off to decrease slightly.
The CdZnTe substrate on this device has not been thinned, nor has an anti-reflection coating been applied to the surface, implying QE$>$0.9 for science-grade devices. CdZnTe thinning should further extend the short-wavelength roll-off to below 1\,\um.

\subsection{Linearity}
\label{sec:linearity}

\begin{figure*}[!t]
  \begin{center}
  \includegraphics[width=1\textwidth]{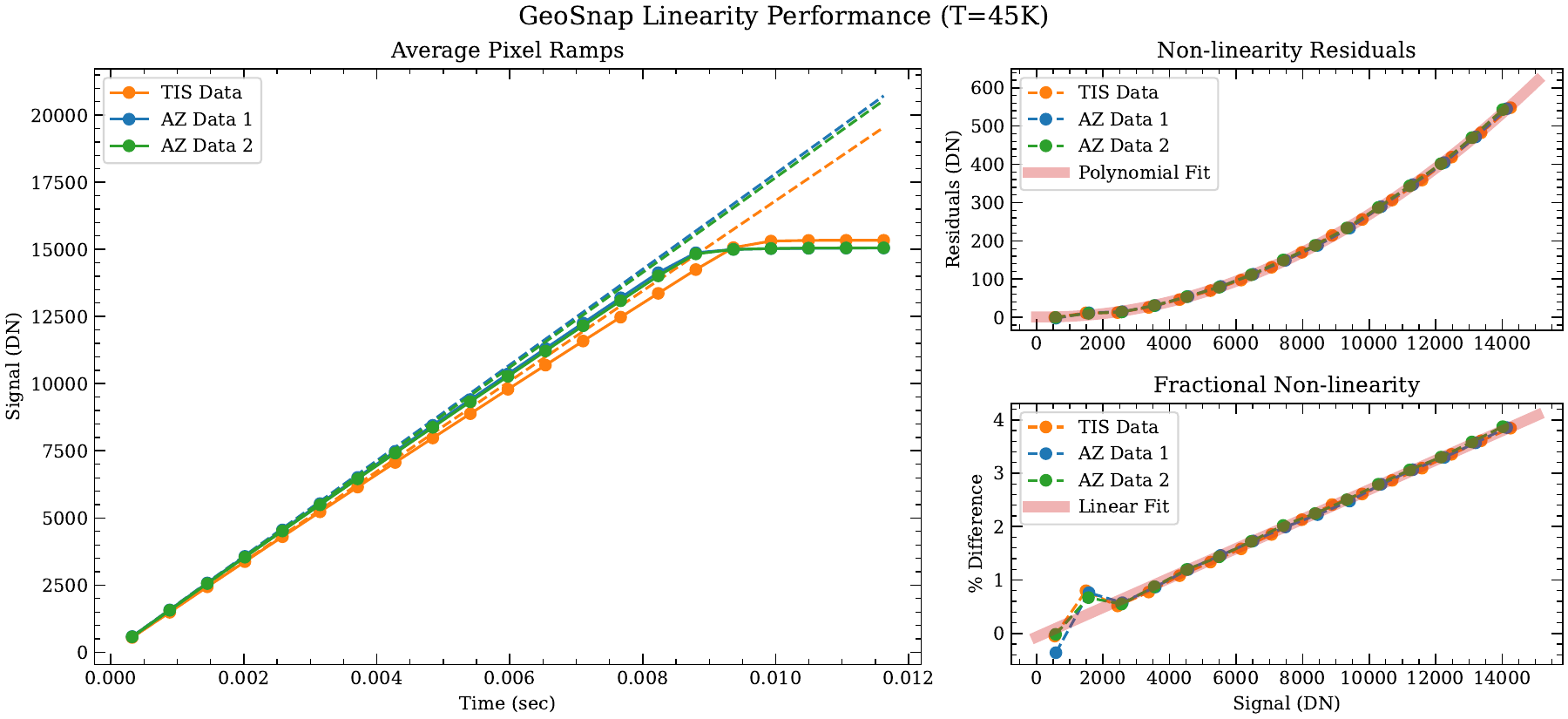}
  \end{center}
  \caption{Linearity data acquired at 85\,Hz frame rate with an operating temperature of 45\,K for three independent linearity runs. \textit{Left:} Average signal for successive integration settings corresponding to intervals of 5\% of the total frame time; the two AZ overlap. \textit{Upper right:} the residuals relative to the linear component of a polynomial fit. These are well fit with a polynomial that can be used to calculate expected residuals based on measured signal level. \textit{Lower right:} Similar to above, fractional non-linearity is linear with signal value. 
  \label{fig:linearity}}
\end{figure*}

Three sets of linearity data were acquired at a temperature of 45K while operating the detector with a frame rate of 85\,Hz: 1) at Teledyne with all four quadrants powered, 2) at Arizona with all four quadrants, 3) again at Arizona with the three bare ROIC quadrants powered down.
For each experimental setup, the input flux was held at a constant rate while changing the fractional integration time per frame, stepping in 5\% intervals to increase the measured signal until saturation.
At each step, 128 frames were acquired and averaged together.

Data were processed by first stacking the averaged frames into a single ``ramp'' data cube. 
A polynomial fit was subsequently performed on each pixel to calculate the zero-flux bias image for each of the three independent datasets. 
The detector bias offsets were then subtracted from their respective ramps.

Prior to testing, we tuned the FPM's internal biases and register to operate at low temperatures from 35\,K to 55\,K. Part of this process included configuring the ADC gain to the recommended $\sim$135\,{\textmugreek}V/DN such that the full range of the 14-bit ADC spans the 2.2\,V input range. 
When including an ADC offset to capture minimum integration images near the detector reset level, pixel values become clipped at the top end of the digitizer before reaching natural saturation. 
Consequently, approximately 67\% of pixels in saturated frames are railed at a value of 16,383\,DN.

\Cref{fig:linearity} shows the median ramp data for each run as well as the small deviations from linearity as a function of signal level. Overall, the ramp data appear to be highly linear across the entire dynamic range with residuals of only a few percent. 
The fractional non-linearity with respect to signal level (bottom right panel of \Cref{fig:linearity}) is well-fit by a linear function with minor deviations at relative low signal levels ($<$2500\,DN). 
These deviations are shown to be repeatable between runs. 


\subsection{Gain \& Well Depth}
\label{sec:gain_well}

\begin{figure}[!b]
  \begin{center}
  \includegraphics[width=78mm]{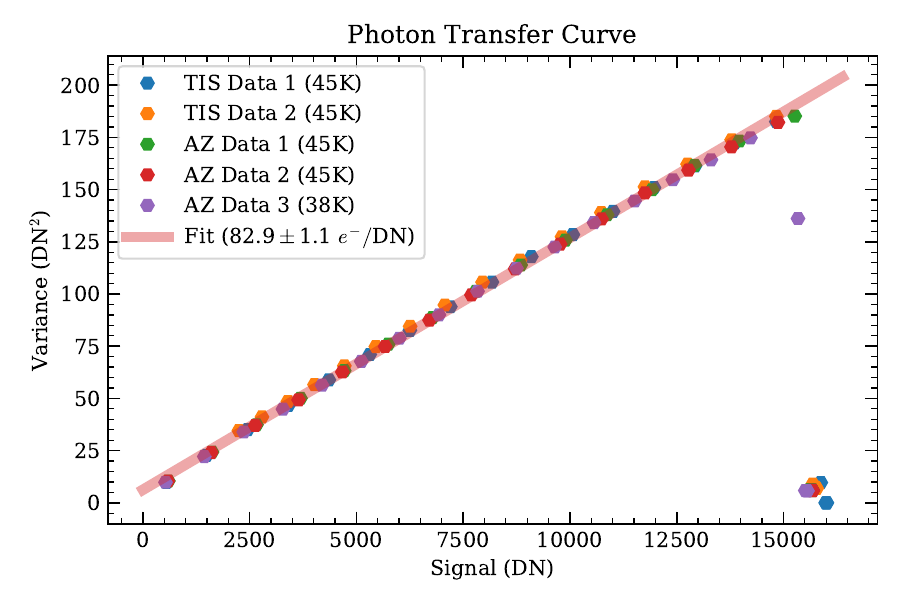}
  \end{center}
  \caption{Resulting GeoSnap photon transfer curve for multiple datasets acquired in a variety of configurations and temperatures.
  \label{fig:ptc}}
\end{figure}

\begin{figure}[!b]
  \begin{center}
  \includegraphics[width=78mm]{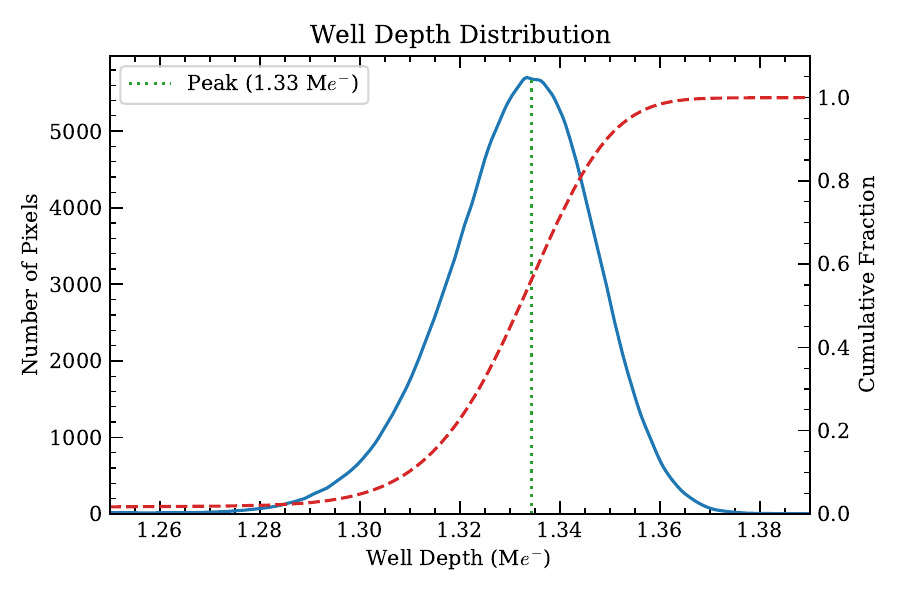}
  \end{center}
  \caption{Histogram of well depths for all pixels hybridized quadrant of the GeoSnap device. The distribution peaks slightly above 1.3\,M\elec. 
  \label{fig:well}}
\end{figure}

\begin{figure*}[!b]
  \begin{center}
  \includegraphics[width=1\textwidth]{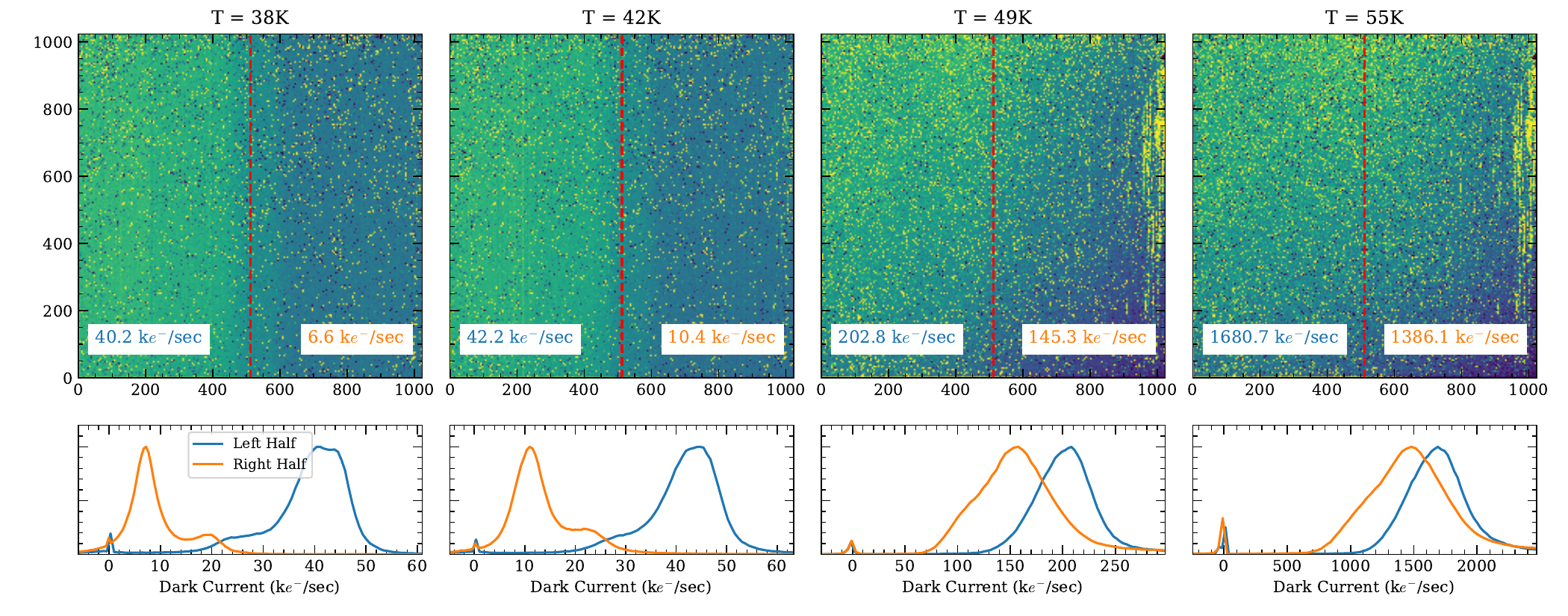}
  \end{center}
  \caption{A sample of dark current slope images acquired at 10Hz framerate in the MITTEN cryostat, demonstrating the evolution of background rates for increasing temperature. Lower figures are histograms of the left and right halves of each slope image. For lower temperatures, there is a clear difference between the two halves, which converges as temperature increases.
  \label{fig:dark_images}}
\end{figure*}

Similar to the linearity data, a series of 128 frames were acquired at a given illumination level to measure effective gain. A variety of independent datasets were acquired at Teledyne and with the MIRAC-5 setup at Arizona. In addition to the three described in \Cref{sec:linearity}, we also analyzed a TIS evaluation dataset that held the fractional integration time constant while changing the total flux from an input blackbody device. In addition, we included a dataset similar to those in the previous section, but acquired at 38K with only the hybridized quadrant powered.

For a given illumination level, we calculated the average signal and variance for each pixel. After applying linearity corrections, the median values of these signal and variance images are plotted to generate the photon transfer curve (PTC) shown in \Cref{fig:ptc}. The PTC curves for all datasets show significant consistency, indicating excellent repeatability between cooldowns. 
IPC corrections were not applied since this effect is absent from the GeoSnap device (\Cref{sec:ipc}).

We adopt gain of 83\,\elec/DN, implying a median well depth of 1.3\,M\elec/pixel for this device (\Cref{fig:well}). 
Because pixel values get clipped at the top end of the ADC range before reaching saturation, the full well depth is likely slightly higher than measured. That is, reducing the DC offset would likely produce a wider sampling of the intrinsic well within the ADC's dynamic range.
The Normal distribution depicted in \Cref{fig:well} is due to pixel-to-pixel offsets in the frame bias, whereas raw data of a saturated frame would show a sharp cut-off at the equivalent of 16,383\,DN.  
About 98.7\% of pixels have well depths greater than 1.2\,M\elec.

\subsection{Dark Current}
\label{sec:bg_rates}

Dark current datasets were acquired with the GeoSnap detector in the MITTEN cryosat at Michigan, which regularly achieved an ambient background temperature of $\lesssim$20\spaceK. 
The combination of low radiative background and absence of optical input ports provides a true measure of the FPM's self-generated background flux.

Detector operating temperatures ranged between 38\,K and 55\,K, where minimum temperature was limited by self-heating of the FPM during operations (0.5\dshrng1\,W of power, depending on operation mode). To minimize the effects of detector power output, data were acquired with a nominal frame rate of 10\,Hz with a single hybridized quadrant. Operating at 10\,Hz also provided the benefit of maximizing the photon collection time for sufficient signal-to-noise. 

For a given temperature setting, the fractional integration parameter, $t_{frac}$, was stepped over a range of 0 to 1 in 0.05 increments. At each step, 128 contiguous frames were recorded and median combined after correcting for residual bias offsets between frames. All pixels within the resulting ramp data cube were fit with a linear function to produce a dark current slope image, one for each temperature setting (\Cref{fig:dark_images}).

The resulting slope images exhibit two flux components roughly split between the left and right halves of the array, particularly evident at the lower temperatures. 
It seems unlikely that intrinsic dark current alone contributes to the bimodal distribution, which is instead likely caused by photon sources internal the FPM.
Incidentally, these two halves correspond to separate output data streams with independent preamps and ADCs. 
The two distributions converge as the operating temperature increases, suggesting detector dark current becomes more prominent. 
We measure a minimum dark current rate of $\sim$6.6\,k\elec/sec, which corresponds to the peak of the distribution of the right side of the 38\spaceK\ slope image.
The images in \Cref{fig:dark_images} also demonstrates an increase in the number of high dark current pixels with temperatures.  

This GeoSnap device does not exhibit the crosshatching pattern attributed to the growth process of the HgCdTe crystal lattice structure because of the applied low bias while operating at a temperature where thermal dark currents dominate. In general, the crosshatching pattern becomes apparent for low thermal dark currents along with an applied bias sufficiently high enough to produce significant tunneling dark current in pixels with defects associated with the lattice mismatch. Such patterns were shown in the LW13 H1RG arrays characterized in \cite{cabr2019} and have been reported in a number of other publications \citep[e.g.;][]{mart2001, chan2008, schl2021}. 

\subsubsection{Dark Current Models}
\label{sec:dark_models}

\begin{figure}[!b]
  \begin{center}
    \includegraphics[width=85mm]{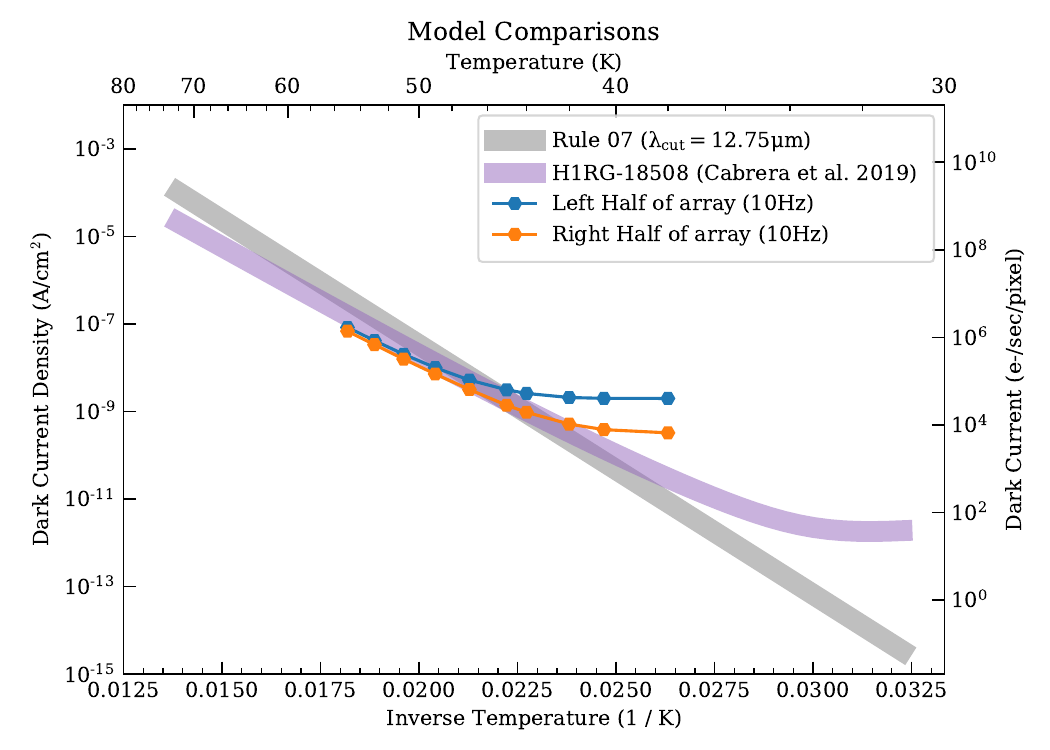}
  \end{center}
  \caption{Arrhenius plot of measured GeoSnap dark current densities at 130mV reverse bias compared to ``Rule 07'' \citep{tenn2008, tenn2010} and H1RG-18508 model fit \citep[Figure 18 of][]{cabr2019}. GeoSnap values at T$>$45\spaceK\ are consistent with those measured for the H1RG part, with values at lower temperatures flattening due to background radiation (e.g., MUX glow).
  \label{fig:dark_models}}
\end{figure}

Dark current behavior for HgCdTe arrays is parameterized in \cite{tenn2008} through the empirically-derived ``Rule~07'' formula, which can be used to generally predict dark current performance for this material based solely on the cut-off wavelength and operating temperature. 
While Rule~07 predictions hold for a range of temperatures and cut-off wavelengths, the performance can be exceeded by optimizing of the fabrication process and device structure.
We plot the the Rule~07 relationship in \Cref{fig:dark_models} assuming a 12.75-\um\ cut-off corresponding to the GeoSnap device. 
At temperatures $>$45\,K, our measured dark current falls below this relationship.

In addition, \Cref{fig:dark_models} shows the dark current model derived from the H1RG-18508 array data presented in \cite{cabr2019}, which incorporates die material originating from the same HgCdTe wafer as this GeoSnap device. 
The dark current model incorporates three primary mechanisms: band-to-band tunneling, generation-recombination (GR), and diffusion.
At T$>$35\spaceK, diffusion and GR dominate the H1RG data, while band-to-band tunneling is the primary contributor at T$<$35\spaceK.

The GeoSnap measurements follow the H1RG-based model fairly well for temperatures greater than 45K. 
The primary deviation occurs at the inflection point where the GeoSnap dark current behavior flattens at lower temperatures, potentially attributed to different operating modes or additional sources of background signal. 
For instance, the H1RG operated with reverse bias of 275\,mV compared to 130\,mV for the GeoSnap, and the GeoSnap utilizes a much higher readout speed with on-chip ADC and consequently larger power dissipation.

Because diffusion and GR dark currents do not change considerably with bias in this regime, good agreement between the GeoSnap and H1RG data at T$>$45\spaceK\ suggests consistent characteristics between detectors for these two contributors. 
In addition, current from band-to-band tunneling (dominant at lower temperatures) is expected to drop for GeoSnap's lower reverse bias, implying that the observed flattening to the GeoSnap curves at T$\sim$45\spaceK\ arises from a different source.
Combined with the distinct differences between the left and right half of the GeoSnap array, these deviations are likely explained by the FPM power dissipation, or multiplexer (MUX) glow. 
The GeoSnap's HgCdTe material therefore appears to have intrinsic dark current consistent with the model derived for the H1RG-18508, which is expected given they are two of the four die from the same growth wafer.

\subsubsection{MUX Glow}
\label{sec:mux_glow}

Measurements of dark current typically include additional background emission as the two are not easily distinguishable. 
In particular, mid- and long-wave detectors are susceptible to self-generated emission \citep{tam1984}, which can be significant compared to the actual dark current of the photosensitive material. 
Previous studies of HxRG detectors have attributed anomalously high dark current to thermal excitation within the multiplexer unit cell rather than a process intrinsic to HgCdTe layer \citep{cabr2019}. 

For the 5-\um\ detectors aboard JWST, this ``MUX glow'' relates directly to the current supplied to the pixel source follower \citep{rega2020}. 
Increasing the current induces local heating within a pixel's unit cell during the sampling process, which is then measured as infrared radiation, especially by material with longer wavelength cut-off.
Further, operating these detectors in subarray witnessed an increase in the glow due to a given pixel being addressed at a higher frequency. That is, sampling the pixel more often accumulates additional glow before prior local heating is able to fully dissipate. 
Compared to H1RG and H2RG data, we suspect that the GeoSnap's measured dark current floor at lower temperatures is MUX glow due to the large on-chip power consumption.

\subsection{Read Noise}
\label{sec:noise}

In this section, we report the read noise for the GeoSnap's standard single-frame readout. 
The GeoSnap ROIC provides an on-chip CDS subtraction that occurs within the normal pixel sampling process, reducing the kTC reset noise contributions by up to 90\% (private communication, TIS). We assess the read noise performance as a function of operating temperature and frame rate.

To compute the read noise, 128 consecutive frames were acquired under dark conditions for a variety of reset times, frame rates, and temperatures in low gain mode. 
The RMS noise per pixel was calculated for each configuration. 
Due to relatively large background signals relative to intrinsic read noise, the photon signal variance was estimated from the frame averages, then subtracted in quadrature from the measured total noise to obtain the read noise values per pixel.

\begin{figure}[!t]
  \begin{center}
  \includegraphics[width=88mm]{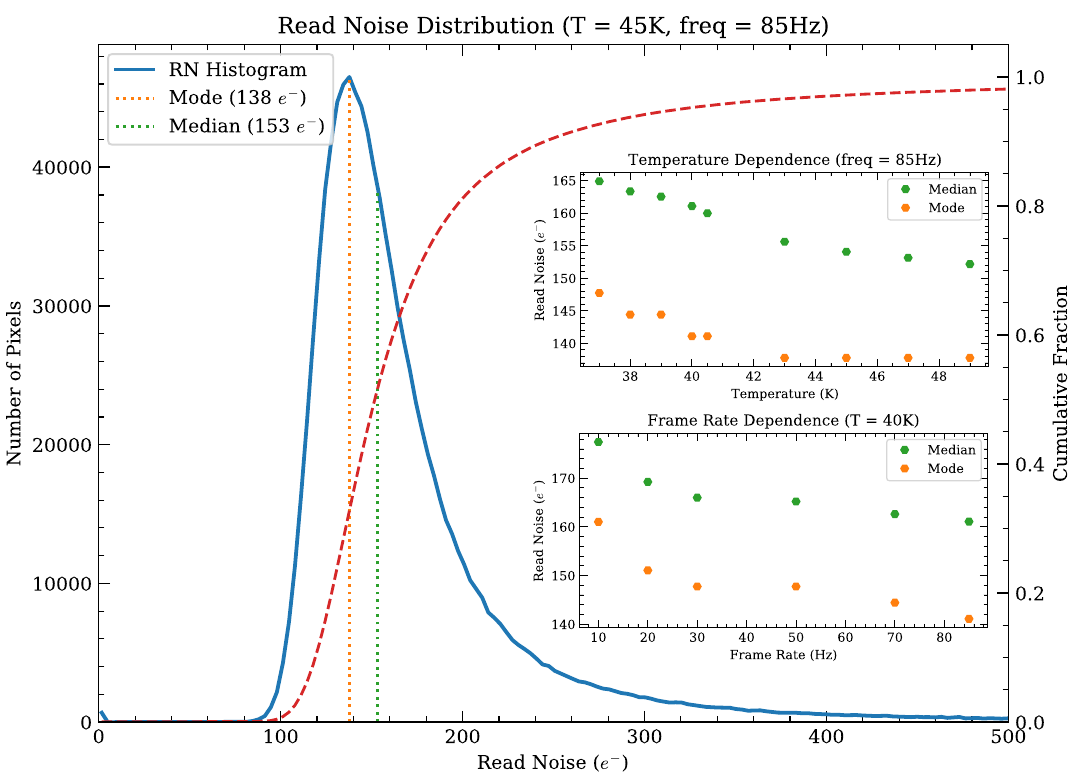}
  \end{center}
  \caption{Detector read noise distribution for typical operating parameters of 85\,Hz frame rate at 45\,K temperature, as a function of temperature (top inset) and frame rate (bottom inset).
  \label{fig:read_noise}}
\end{figure}

The left panel of \Cref{fig:read_noise} shows a histogram of the pixel noise operating at 45\spaceK\ with a frame rate of 85\,Hz.
Read noise for a typical single frame in the low gain mode is measured to be approximately 140\,\elec\ RMS, corresponding to the peak of the histogram, whereas the median is slightly higher ($\sim$150\,\elec) due to the extended tail of the distribution. For this particular device, noise appears to increase slightly as the detector operating temperature decreases, although the change is a mere 7\% over $\sim$10\spaceK. 
The low temperature increase may be related to the photon noise subtraction, over-correcting for larger signal values or under-correcting at lower signal values. 
There is also evidence for temperature-dependent read noise in H2RGs (private communication, JWST NIRCam team). 

Similarly, the read noise is observed to increase towards lower frame rates. 
The rise may be due to larger contributes from \invf\ noise component (\Cref{sec:pink_noise}) for larger sampling intervals.

\begin{figure}[!b]
  \begin{center}
  \includegraphics[trim={10mm 10mm 20mm 20mm},clip,width=88mm]{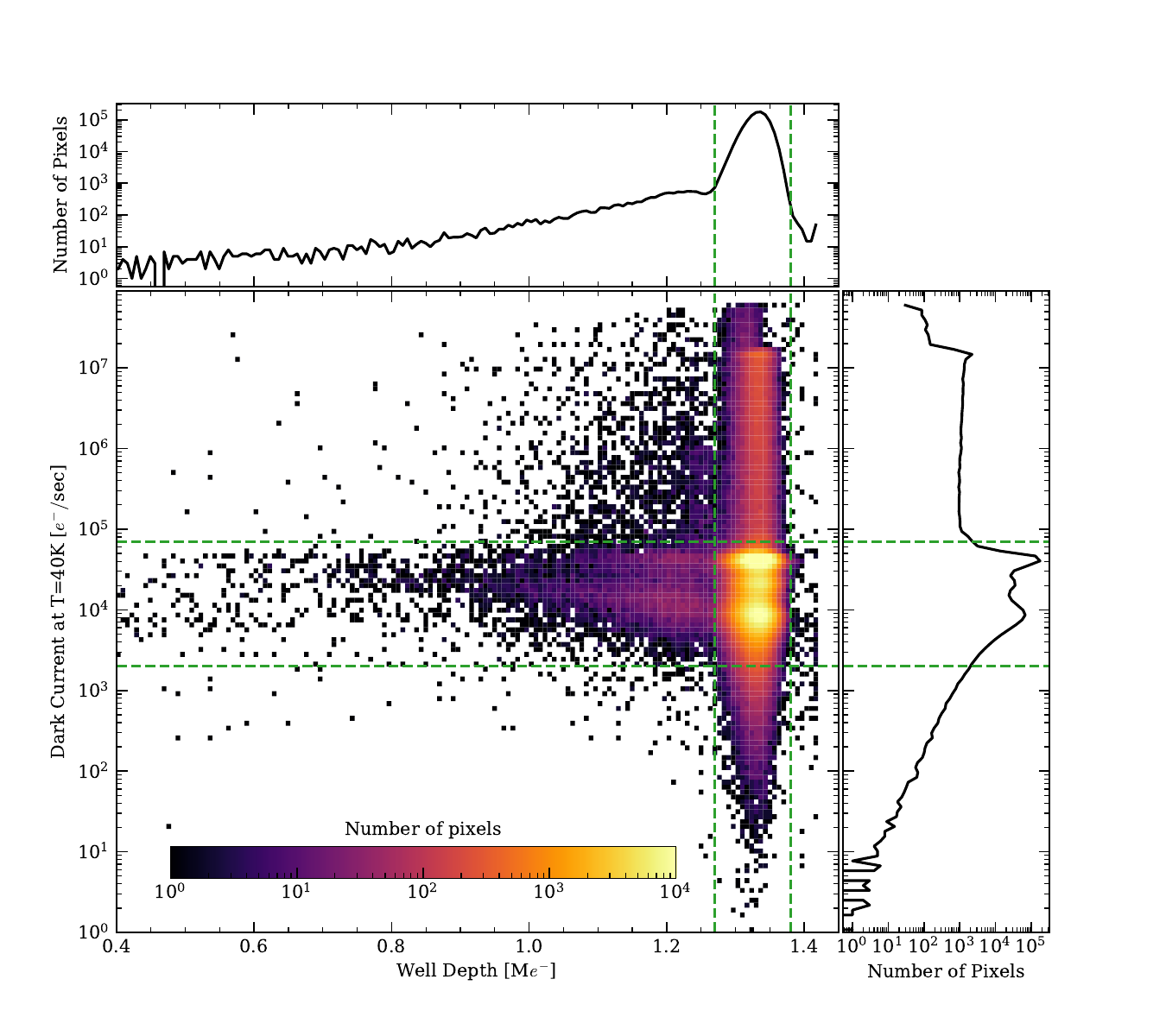}
  \end{center}
  \caption{2D histogram of dark current versus well depth obtained with a focal plane temperature of 40\spaceK. The distribution is centered at 40.4\,k\elec/sec and 1.33\,M\elec. Approximately 88\% of pixels reside within the dashed rectangular region surrounding the peak of the distribution.
  \label{fig:dark_well}}
\end{figure}

\begin{figure*}[!b]
  \begin{center}
    \includegraphics[width=1\textwidth]{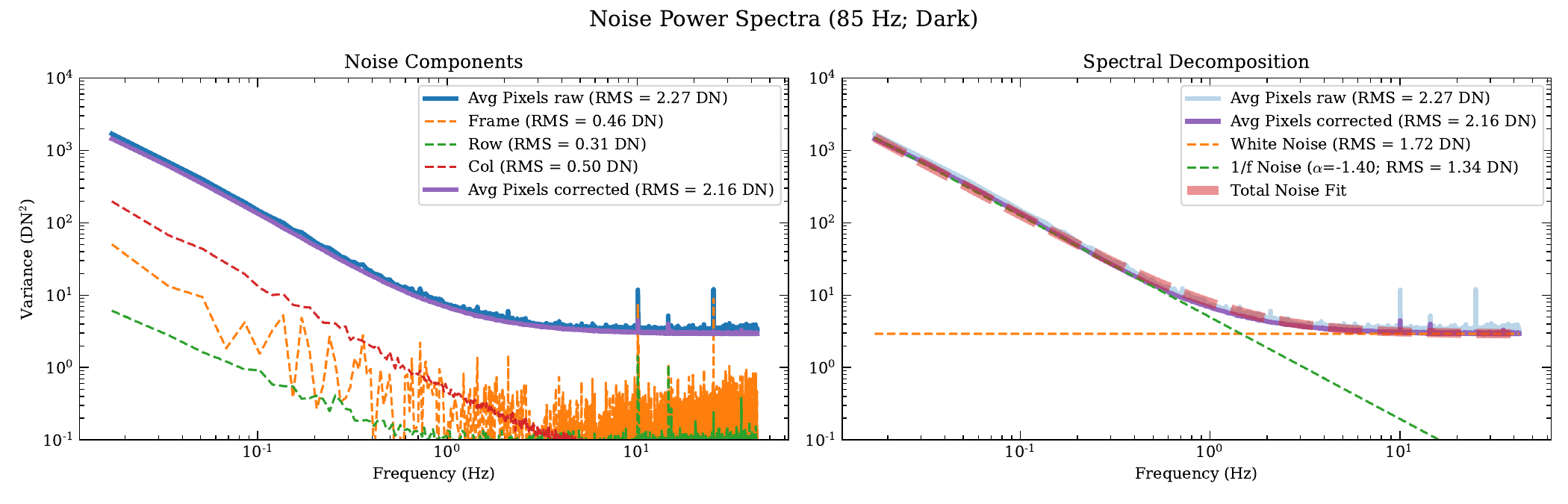}
  \end{center}
  \caption{\textit{Left:} Average pixel noise power spectrum in units of variance (DN$^2$) before and after removal of correlated components (frames, rows, columns) for a sequence of 5000 dark observations. Data were acquired at a frame rate of 85\,Hz and temperature of 40\spaceK. \textit{Right:} Final noise profile fit breaking out white noise and \invf\ noise components. In this case, the \invf\ component has a best-fit exponent of $\alpha=-1.4$.
  \label{fig:3dnoise}}
\end{figure*}

\subsection{Persistence}
\label{sec:latents}

Our test environments did not include a mechanism for effectively measuring latent signals related to detector persistence. Regardless, we were unable to visually identify any obvious evidence of persistence in our data when switching between partially saturated frames and dark frames. This is not unexpected, given the combination of the CTIA pixel cell held at a constant voltage, high-speed readout inhibiting build-up of charge within traps, and the long wavelength cut-off where persistence tends to be reduced compared to shorter wavelength devices \citep{leis2016}.

\subsection{IPC}
\label{sec:ipc}

We used dark data presented in \Cref{sec:bg_rates} to assess the presence of interpixel capacitance \citep[IPC;][]{moor2004}. By selecting isolated hot pixels with excessively high dark current rates, we can determine the coupling between these pixels to their nearest neighbors. We isolated $\sim$8300 hot pixels, removed their surrounding background levels, then normalized the total signal within the 3$\times$3 region surrounding the hot pixels to a value 1. After stacking and median-combining the normalized image, we calculate a coupling constant $\alpha_{ipc} = 0.001 \pm 0.005$, which is consistent with 0. This is expected, because the CTIA readout circuit holds the pixel gates at the same constant voltage, which is predicted to remove the effects of capacitive coupling between neighboring pixels.

\subsection{Operability}

This engineering-grade array shows few cosmetic defects as exhibited the images in \Cref{fig:overview}. Even though, flat field illuminated image shows significant non-uniformity, this pattern is consistent and easily removed through standard data reduction processes.

We calculate the fraction of operable pixels by measuring the total number of pixels with acceptable read noise, dark current, and well depths at an operating temperature of 40\spaceK. \Cref{fig:dark_well} shows the distribution of pixels for a histogram image of dark current versus well depth. Following the data presented in \Cref{sec:gain_well,sec:bg_rates}, the distribution is centered near a dark currents of 40\,k\elec/sec and 1.3\,M\elec\ well depth. 
Approximately 95\% of pixels have dark currents below $10^5$\,\elec/sec, well depths of $>$1.25\,M\elec, and read noise values less than 500\,\elec. 
We expect the number of inoperable pixels to increase with temperature, as evidenced in \Cref{fig:dark_images}.

\section{Excess Noise Analysis}
\label{sec:pink_noise}

Excess low frequency noise (ELFN) is a property of previous generation of Si:Sb Blocked Impurity Band \citep[BIB;][]{stap1984} and Si:As Impurity Band Conduction \citep[IBC;][]{arri1998} detectors used for longwave infrared astronomy. ELFN is produced within the detector material, specifically resulting from additional photon absorption within the the IBC blocking layer. This behavior prevents the noise from reducing as the square root of the number of reads, especially in high-flux regimes. This effect has been widely reported in Raytheon AQUARIUS Si:As arrays \citep[e.g.]{ives2012,ives2014,hoff2014}, but should be absent in the GeoSnap's HgCdTe detector material owing to the different material and unit cell architecture. 

As expected, the GeoSnap's pixel noise power spectrum does not exhibit the characteristic ELFN profile observed in the AQUARIUS array. However, the noise power spectrum instead shows a prominent and troubling \invf\ profile (\Cref{fig:3dnoise}). 
We identified multiple sources of \invf\ noise common to the rows, columns, and full array, which are readily subtracted during standard data reduction processes. However, a significant residual \invf\ component remains as it cannot be removed through common mode analysis (e.g., subtraction of neighboring pixels). 
For dark observations operating at 85\,Hz frame rate, the \invf\ component dominates over the underlying Gaussian noise on timescales longer than $\sim$1 sec (frequencies below $\sim$1\,Hz). 
In addition, the slope of the \invf\ component does not always scale exactly with frequency. Instead, its power spectrum has the functional form,
\begin{equation}
  S(f) \propto f^{\alpha}, 
  \label{eq:invf}
\end{equation}
where the exponent, $\alpha$, has a best-fit value that varies between -1 and -1.5.

\begin{figure}[!t]
  \begin{center}
    \includegraphics[width=88mm]{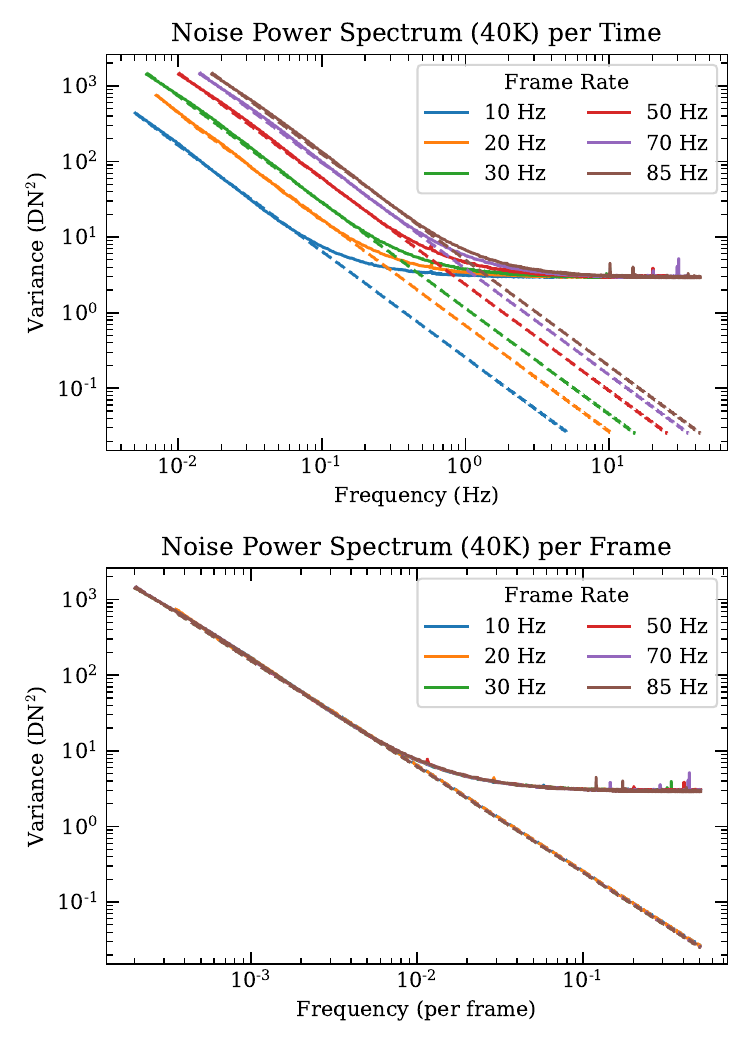}
  \end{center}
  \caption{The noise power spectra of dark observations acquired at different frame rates. The upper panel shows the power spectrum for each data set plotted as a function of temporal frequency, whereas the same data is plotted in the lower panel as a function of "per frame" (ie., dividing the values of each curve by their corresponding frame rate). Dashed lines indicate individual fits to the \invf\ component.
  \label{fig:freq_shift}}
\end{figure}

\subsection{Frame Rate and Integration Time}

We acquired a number of data sets with changing frame rates to assess any modulation of the \invf\ noise.
\Cref{fig:freq_shift} demonstrates that the position of the \invf\ component within the power spectrum directly correlates with the frame rate. That is, the power spectra for different frame rates and their \invf\ components perfectly overlap when plotting with respect to the frequency on a ``per frame'' basis rather than with respect to Hz (lower panel of \Cref{fig:freq_shift}). This suggests that this noise must originate from an internal process that does not depend on frame rate. As a consequence, observations with lower framerates will have effectively less \invf\ noise contributions for a fixed period of time. 

In addition, we checked for dependency of \invf\ profile on the fractional integration time. The GeoSnap provides the ability to control the fraction of a frame time that is dedicated to integration by configuring the duration of the pixel cell reset switch. For both dark and half-well observations, the \invf\ noise profile did not change with respect to the fractional integration time. 

\begin{figure}[!b]
  \begin{center}
    \includegraphics[width=88mm]{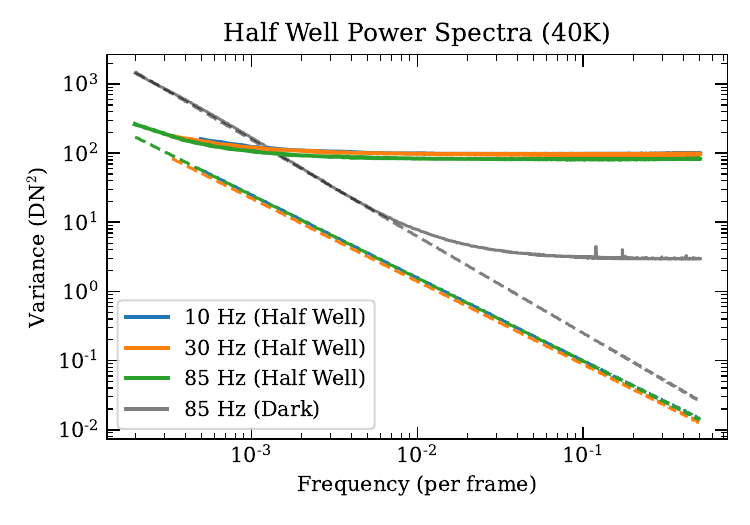}
  \end{center}
  \caption{Comparison of power spectrum observed at 50\% well fill for a range of frame rates plotted as a function of frame frequency, following the lower panel in \Cref{fig:freq_shift}. Dashed lines indicate the fitted \invf\ component. The power spectrum for a dark observation is provided for reference, showing that half-well observations produce significant reduction in \invf\ component both absolutely as well as relative to the white (photon) noise.
  \label{fig:halfwell}}
\end{figure}

\subsection{Signal Level}
\label{sec:signal_level}

This noise profile is also observed within the non-hybridized quadrants of the array with the same characteristics as dark observations in the hybridized quadrant, indicating the \invf\ noise arises somewhere within the pixel cell's readout chain. Unlike the AQUARIUS arrays, this is not a phenomenon related to the HgCdTe detector material, which would have likely been identified in previous HxRG devices.

While this level of noise would be challenging for low-flux applications, the absolute contributions from the \invf\ component appears to lessen with increased signal levels. \Cref{fig:halfwell} shows that the noise properties are significantly improved when the device is uniformly illuminated to a 50\% well level compared to dark observations. The \invf\ noise is reduced and the slope is less steep with a best-fit exponent of $\alpha=-1.2$, suggesting that signal levels somehow affects manifestation of the \invf\ noise. Combined with the overall increase of photon noise, these two effects favorably conspire to push the upturned knee of the observed power spectrum profile to much lower frequencies. Observations with high background rates then become tractable through moderately slow chopping and/or nodding of the telescope, allowing an observer to re-position their source within the detector's FoV on manageable timescales without significant overhead.

\begin{figure}[!t]
  \begin{center}
    \includegraphics[width=88mm]{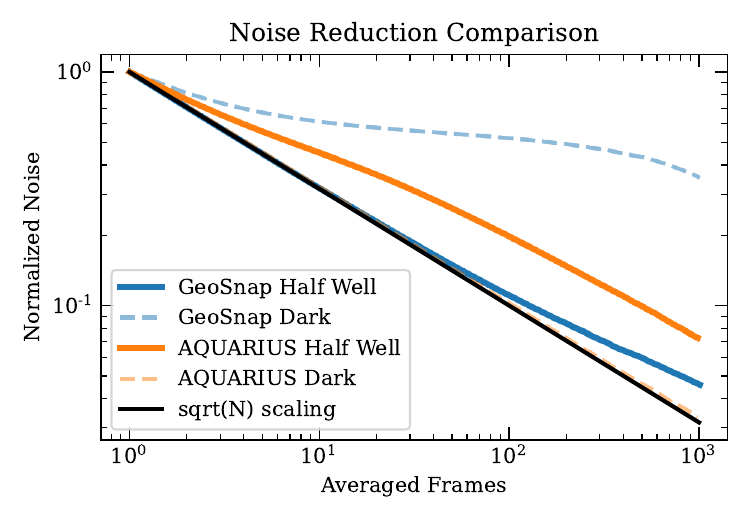}
  \end{center}
  \caption{Noise reduction as a function of number of averaged frames, comparing GeoSnap with LBTI/NOMIC's AQUARIUS data for dark and half well signal scenarios.   
  \label{fig:noise_scaling}}
\end{figure}

\Cref{fig:noise_scaling} compares the noise scaling as a function of consecutively averaged frames for dark and half-well illuminated scenarios for both the GeoSnap and AQUARIUS arrays. The AQUARIUS data were acquired using the LBTI/NOMIC instrument (private communication, S.\ Ertel). For each scenario, we obtained 2000 consecutive frames, subtracted the average relative offsets between frames to remove any electronic bias offsets, and generated bad pixel masks. Each sequence of images was then split into first and second halves (as would be experienced for a telescope nod offset), which were subtracted frame-by-frame, leaving a sequence of noise images. We then successively averaged together the differenced images and measured the resulting noise after applying a bad pixel mask.

In \Cref{fig:noise_scaling}, we recover the ELFN behavior of LBTI/NOMIC's AQUARIUS array, where the noise drop-off for dark observations very closely follows the ideal $\sqrt{N}$ scaling, but higher well depths produce a large noise excess.
GeoSnap's \invf\ behavior is effectively the inverse of ELFN with respect to signal. For the dark observations case, our chosen frame subtraction baseline (1000 frame times) means that the \invf\ contributions become the dominate noise source, preventing efficient reduction of the noise through frame averaging. Instead, the low-signal scenario would require subtraction of frames pairs much closer in time. At high signal levels, though, there is a $\sim$40\% noise excess (compared to 250\% for the AQUARIUS), which only improves with higher well depths; GeoSnap is the more favorable device in high-flux regimes. 

\subsection{Temperatures}

A number of \invf\ noise observations were acquired for a range of temperature between 38\spaceK\ and 52\spaceK\ in both dark and half well configurations. The power spectrum profiles for the two well level settings were consistent across temperatures. We conclude that the amplitude and $\alpha$ value of the \invf\ excess noise does not depend on temperature.

\subsection{Sample Up the Ramp}

\begin{figure}[!b]
  \begin{center}
    \includegraphics[width=88mm]{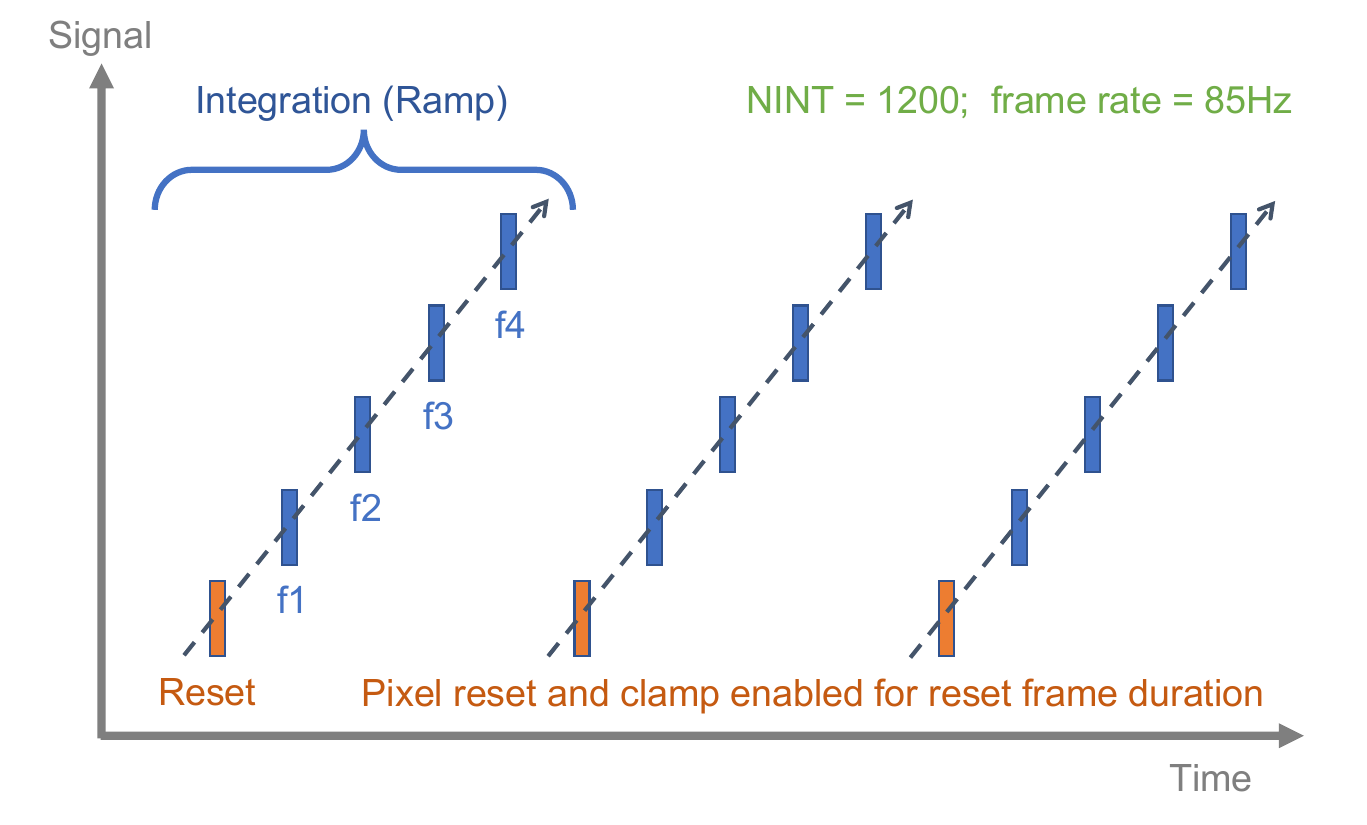}
  \end{center}
  \caption{An illustration of the up-the-ramp readout scheme implemented to investigate \invf\ noise as a function of signal levels. Each integration consists of a single reset frame and four non-destructive read frames. All frames, including the reset frames, are written to disk.   
  \label{fig:sur}}
\end{figure}

In general, \invf\ noise can be conceptualized as a slowly varying pixel offset whose distribution is non-Gaussian and power spectrum follows the form in \Cref{eq:invf}. Sampling this signal multiple times before it significantly changes allows one to effectively remove the \invf\ contribution. That is, subtracting consecutive frames will effectively eliminate the \invf\ component, resulting in primarily white noise residuals. If the source of this noise were simply a slowly varying DC offset, then a readout scheme implementing correlated double sample (CDS) or sample-up-the-ramp (SUR) readout should remove it, assuming sampling is fast enough.

We implemented a custom firmware to acquire sample-up-the-ramp data (\Cref{fig:sur}). This mode deactivates the internal on-chip CDS subtraction, holds the pixel reset switch closed on the ``0th'' frame, and then non-destructively reads some number of frames as the on-chip capacitor continues to accumulate charge until the next reset frame. Our ramp integrations consisted of one reset frame and four read frames, which are written to disk as a separate file per frame. Data were acquired at 85Hz with 1200 integrations (e.g., 6000 total frame). The average well level in the final read frame were approximately 90\%. As a control, we simultaneously acquired MUX data from a bare quadrant when evaluating the behavior of the hybridized quadrant.

\begin{figure}[!t]
  \begin{center}
    \begin{minipage}{\columnwidth}
      \includegraphics[width=88mm]{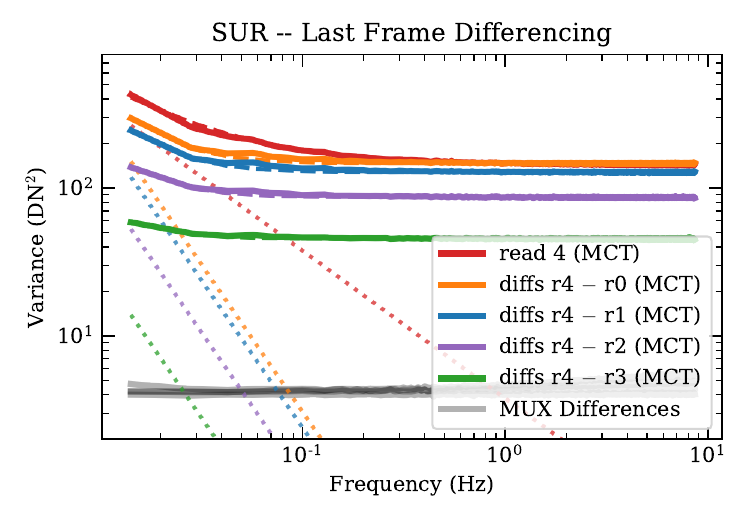}
    \end{minipage}
    \begin{minipage}{\columnwidth}
      \includegraphics[width=88mm]{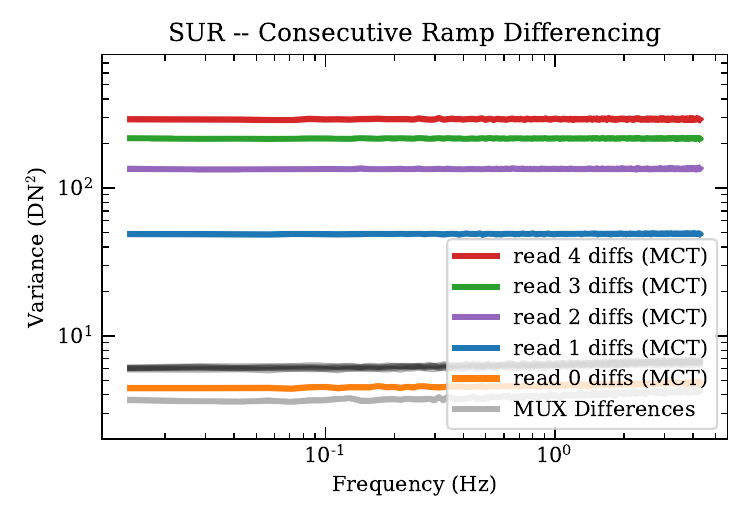}
    \end{minipage}
  \end{center}
  \caption{\textit{Upper:} Power spectra for SUR data processed in various manners. The colored lines indicate subtraction of some frame within the ramps from the final read frame. Dashed lines are the fitted \invf\ components. Frames pairs with larger signals differences reveal larger residual \invf\ noise after subtraction. Whereas the MUX data acquired at the same time (in an non-hybridized quadrant with no charge accumulation) show complete removal of the \invf\ component, presumably because the data have the same signal levels.
  \textit{Lower:} Differences of the same read frames in consecutive integrations. Since all images for the same read frame have similar signal levels, the \invf\ component is always removed.
  \label{fig:sur_diffs}}
\end{figure}

Under dark conditions, performing ramp fitting as well as last-minus-first read subtraction completely removed the 1/f component in both the hybridized and ROIC quadrants (\Cref{fig:sur_diffs}). While encouraging, results from \Cref{sec:signal_level} indicated that \invf\ noise profile can be affected by the accumulated signal. 
Indeed, when illuminating the detector, a significant residual \invf\ component exists after performing intra-ramp frame subtraction. The amount of residual \invf\ directly relates to initial signal differences between the frames. For instance, $R_4 - R_3$ performs better in this regards than $R_4 - R_1$. The paradigm of a single \invf\ signal propagating through consecutive reads appears to be incomplete. Instead, the \invf\ component appears to be modulated by the accumulated charge in some manner. This is an active area research by Teledyne to isolate the source and mitigate this electronic noise.

To further highlight this signal dependency, subtracting the same read frame in subsequent ramps completely removes the \invf noise component; the power spectra are perfectly flat, consistent with Gaussian noise (\Cref{fig:sur_diffs}, lower panel). We additionally attempted subtracting various combinations of frame pairs, but the only scenarios that completely removed the \invf\ noise component required equal signal levels for the non-consecutive subtraction pairs.

\subsection{Mitigation Strategies}

In order to better isolate the source of the \invf\ noise within the readout chain, we generated a few data sets that turned off various components within the unit cell. So far, we have shown that the noise cannot originate from the HgCdTe material (present in both MUX and active pixels) or the CDS functionality (CDS was turned off during the ramp sampling procedure). We further attempted to isolate the CTIA by always keeping the CDS clamp switch closed as well as setting the CTIA bias to 0\,nA. In both cases, the \invf\ noise was still present at the same levels. Hence, it appears that the noise does not originate in the CTIA circuit, but must originate in components after the CTIA and prior to (or during) A/D conversion. 

The latest \texttt{B0} ROIC revision qualitatively shows the same \invf\ noise contamination (private communication, D. Ives), meaning that instruments employing these devices require mitigation strategies until such time that the source of the noise is isolated and removed in future firmware or hardware updates.

We have identified a number of observing strategies that will help reduce or \invf\ noise contributions:
\begin{enumerate}
   \item Operate close to full well. For high-background observations, the new ROICs come equipped with larger well depths of $\sim$2.4\dshrng2.6\,M\elec\, allowing photon noise to dominate over the \invf\ noise component (\Cref{fig:halfwell}). 
   \item Operate at lower frequencies. Since the \invf\ noise profile is generated per frame rather than per Hz, operating at lower frame rates effectively shifts \invf\ noise to lower frequencies and reduces, allowing more efficient nodding (\Cref{fig:freq_shift}).
   \item Perform chopping and/or nodding. The \invf\ noise can be effectively removed by subtracting pairs of observations where the source has been modulated via tip/tilt offsets faster than the characteristic frequency corresponding to the up-turn of the \invf\ noise component. This strategy will be implemented in the ELT/METIS instrument \citep{paal2014, bran2022}.
   \item For low-background observations, a CDS or SUR mode may provide an effective way to subtraction out the \invf\ noise if signal accumulation is low. This work did not investigate GeoSnap's high-gain mode, which may be more suitable for such a scenario. 
\end{enumerate}

\section{Conclusions}\label{sec:conclusions}

We performed testing and characterization of the first mid-IR GeoSnap-18 device produced specifically for ground-based astronomy. This engineering-grade part consisted of an early \texttt{A0} ROIC hybridized with a 13-\um\ HgCdTe die onto a single 1024$\times$1024 quadrant. The photo-sensitive region was not AR-coated and retained its CdZnTe substrate, limiting QE to $\lesssim$70\%. 

The device was found to have the following properties while operating in low gain mode:
\begin{itemize}
  \item The GeoSnap showed a well depth of $\sim$1.3\,M\elec\ with a gain of 83\,\elec/DN. We were unable to probe the full range of the well depth due to hitting the upper limit of the ADC.
  \item The pixel response showed a well-behaved non-linearity with a $\sim$4\% maximum deviation from linear at the highest signal levels.
  \item Dark current measurements at T$>$45\,K were consistent with HgCdTe models derived for H1RG detectors bonded to similar HgCdTe material. Measurements below 45\,K were limited by MUX glow, flatting out with a lower limit of 6,600\,\elec/sec.
  \item The detector read noise is measured to be approximately 140\,\elec\ RMS, corresponding to the peak of the histogram, whereas the median is slightly higher ($\sim$150\,\elec) due to distribution's extended tail.
  \item Approximately 95\% of pixels have dark currents below $10^5$\,\elec/sec, well depths of $>$1.25\,M\elec, and read noise values less than 500\,\elec. 
  \item We did not find any evidence for persistence or IPC in the gathered GeoSnap data. 
\end{itemize}

While we confirmed the GeoSnap array is free of ELFN, it instead suffers from a \invf\-like noise originating within the pixel readout. This noise severely limits sensitivity in low-signal regimes and can dominate on timescales greater than a couple seconds for the fastest frame rates. We observed the following behaviors:
\begin{itemize}
  \item The power spectrum for dark observations operating at 85\,Hz frame rate were dominated by \invf\ component on time scales greater than $\sim$1 sec.
  \item The \invf\ power spectrum profile was observed to shift with frame rate such that \invf\ components always exhibited the same power spectrum when plotted with respect to a ``per frame`` frequency.
  \item Observations acquired at half well showed a lower overall \invf\ contribution along with a more shallow slope compared to dark observations.
  \item Analysis of sample-up-the-ramp observations showed that a residual \invf\ contribution remained after subtracting consecutive frames with different signal levels. However, subtracting frames with the same accumulated signal from consecutive ramp integrations completely removed the \invf\ component. These results suggest the \invf\ component is modified by signal level.  
  \item GeoSnap's detector noise averages down much more favorably at high signal levels compared to the AQUARIUS array.
  \item We had determined that the source of the \invf\ noise does not originate within the CTIA or CDS circuit, but must occur prior to or during the A/D conversion.
\end{itemize}

There are a few possible methods to mitigate the effect of the \invf\ noise:
\begin{enumerate}
  \item implement fast chopping to modulate the location of an astronomical source on the detector.
  \item reduce the speed of the detector frame rate to shift the \invf\ power spectrum to lower frequencies.
  \item operate in a high-background environment where photon noise dominates over \invf\ noise. 
\end{enumerate}

Despite the engineering-grade qualities of the device presented in this work, this detector is well-suited to perform high-quality mid-IR observations with ground-based instruments in high-background regime. The large well depths and high-speed frame rate provide significant advantages over HxRG arrays with similar cut-off wavelengths, while the noise properties have favorable performance relative to the Raytheon AQUARIUS array especially for high signal levels. Near-term work includes commissioning the GeoSnap array mounted in the MIRAC-5 camera while deployed behind the MAPS AO system on the MMT.

New science grade devices will take advantage of double the well depths in the \texttt{B0} ROICs as well as higher overall QE from AR-coated and CdZnTe-thinned surfaces \citep[e.g., METIS;][]{bran2022}. Future design and engineering work is needed to isolate the origin of the \invf\ noise as well as mitigate the MUX glow limiting the background dark rate.


\section*{Acknowledgments}
We thank the anonymous referee for their review, which has improved the quality of the paper and helped clarify several details.
This work was supported through funding by the \fundingAgency{Templeton World Charity Foundation} under grant \fundingNumber{TWCF0330} and \fundingAgency{Heising-Simons Foundation} under grant \fundingNumber{2020-1699}.
We thank Dr.\ Steve Ertel and the LBTI/NOMIC team for loan of their AQUARIUS noise data, and are grateful for the engineering staff at Steward Observatory, The University of Arizona including Manny Montoya, Grant West, and Dennis Hart. 
We are also grateful for the collaboration of our colleagues at the University of Michigan Department of Physics, in particular Dr.\ John Monnier, Megan Morgenstern, and Paul Thurmond and his team in Randall Laboratory, the Space Physics Research Lab, and the Plasmadynamic and Electric Propulsion Lab.




%

\bibliography{geosnap}%

\end{document}